\documentclass{vhepu14}

\def\Journal#1#2#3#4{{#1} {\bf #2}, #3 (#4)}

\def\PLB{{\em Phys. Lett.}  B}
\def\PRL{\em Phys. Rev. Lett.}

\def\be{\begin{equation}}
\def\ee{\end{equation}}
\def\bea{\begin{eqnarray}}
\def\eea{\end{eqnarray}}

\newcommand{\Photo}{}

\begin{document}
\vspace*{4cm}

\title{From Microscopic Gravitational Waves to the Quantum Indeterminism}

\author{ Vo Van Thuan }

\address{Vietnam Atomic Energy Institute (VINATOM)
\\and
\\NEPIO Office
\\59 Ly Thuong Kiet street, Hoan Kiem district, Hanoi, Vietnam
\\Email: vvthuan@vinatom.gov.vn}

\maketitle\abstracts{
Based on an extended space-time symmetry a new attempt to search for links between general relativity and quantum mechanics is proposed. A simplified cylindrical model of gravitational geometrical dynamics leads to a microscopic geodesic description of strongly curved extradimensional space-time which implies a duality between an emission law of microscopic gravitational waves and the quantum mechanical equations of free elementary particles. Consequently, the Heisenberg indeterminism would originate from the time-space curvatures.}

\section{Introduction}

A fundamental problem of physics is the consistency of quantum mechanics with general relativity. Kaluza and Klein ~\cite{Ka1}$^{,}$~\cite{Kl1} were pioneers to propose a space-like extradimension (ED) which is to compact in a micro circle in a relation to general relativity. Klein and Fock ~\cite{Kl1}$^{,}$~\cite{Fo1} also found a formalism that the equation of motion of a massive particle in 4D space-time can be obtained by reducing the EDs of a massless particle in a higher dimensional time-space. For the semi-classical approach to quantum mechanics introduced by de Broglie and Bohm ~\cite{Br1}$^{,}$~\cite{Bo1} the hidden parameters are somehow reminiscent of EDs. Later on, the evidence for violation of Bell inequalities ~\cite{Be1}$^{,}$~\cite{Fr1} abandoned the models with local hidden parameters, however, leaving the door open to non-local hidden variables. In the wake of high dimensional superstring models, another trend has been developed following the Kaluza-Klein geometrical dynamics, of which most applied space-like EDs, while few others considered time-like ones. There are two main approaches with time-like EDs: membrane models in the Anti-de-Sitter geometry (AdS), such as ~\cite{Ma1}$^{,}$~\cite{Ra1}and induced matter models~\cite{We1}$^{,}$~\cite{We2}. In particular, Maldacena ~\cite{Ma1} found a duality between AdS and conformal fields as AdS/CFT formalism. Randall and Sundrum ~\cite{Ra1} applied an infinite 5D AdS model for a hierarchy solution. For the induced matter approach, Wesson~\cite{We1} has proposed a space-time-matter model describing proper mass as a time-like ED. A geometrical dynamic model for elementary particles was proposed by Koch~\cite{Ko1}$^{,}$~\cite{Ko2} with a time-like ED which offered a method for derivation of Klein-Gordon equation. Our preliminary study~\cite{Vo1} following the induced-matter approach was based on the space-time symmetry in which the Klein-Fock reduction formalism was used. In the present study we do a new attempt in a more direct way to prove that the quantum wave equations in 4D space-time can be identical to a general relativistic geodesic description of curved extradimensional time-space.

\section{Symmetry with time-like extradimensions}

Considering two orthogonal subspaces, 3D-time and 3D-space, we construct an ideal 6D flat extended symmetrical time-space $\{t_1,t_2,t_3 \mid x_1,x_2,x_3\}$:
 \begin{equation}
dS^2=d\vec{k}^2-d\vec{l}^2=dt_k^2-dx_l^2;
\label{eq1}
\end{equation}
 Our further investigation bases on the time-space symmetrical "lightcone" embedded in the 6D flat time-space $(\ref{eq1})$:

\begin{equation}
d\vec{k}^2=d\vec{l}^2\qquad \rightarrow\qquad dt_k^2=dx_l^2;
\label{eq2}
\end{equation}

Where $k,l=1 \div 3$ are summation indexes that: $d\vec{k}^2={\sum dt_k}^2$ and $d\vec{l}^2={\sum dx_l}^2$. Hereafter, natural units ($\hbar=c=1$) are used generally unless it needs an explicit quantum dimension. Any differentials-displacements of $dt_k$  and $dx_l$ correspondingly in 3D-time and in 3D-space can be independent from each other. They are naturally considered as the most primitive sources of physical potential energy in the symmetrical time-space. The time-like source can form a special global vacuum potential in 3D-time equivalent to the original tachyonic Higgs potential. Another half of the primitive source in 3D-space would contribute to the global cosmological constant $\Lambda$. We assume a postulate that the equality $(\ref{eq2})$  between the squared linear time-like and space-like intervals $d\vec{k}^2=d\vec{l}^2$ is to accept as a fundamental time-space symmetry not only for Euclidean geometry, but also to be conserved for the linear translational elements of more generalized space-time geometries and in their transformation from a higher dimensional geometry to lower one. This assumption denoted as a principal conservation of linear translation (CLT) bases on our experience of the firm Lorentz invariance as well as of the homogeneity and isotropy of 4D Minkowski space-time.
\\Now let us consider among differentials $(\ref{eq2})$ those displacements for which a time-like displacement correlates with a space-like one by a harmonic function $f(t_k,x_l)$.  In the flat 6D time-space we introduce the following 6D isotropic plane wave equation:

\begin{equation}
\frac{\partial^2f }{\partial t_k^2}=\frac{\partial^2f }{\partial x_l^2};
\label{eq3}
\end{equation}

Where $t_k$  and $x_l$ remain Descartes coordinates, describing transmission of plane waves $f(t_k,x_l)$ in Euclidean time-space. The 6D wave transmission $(\ref{eq3})$ can serve as a primitive energy-momentum formation of physical objects, symmetrical in 3D-time and 3D-space. Most of the primary form of energy in 3D-time probably is almost unobservable from our 3D-space and would be a kind of dark energy. Zeldovich in~\cite{Ze1} applied a special global cosmological vacuum to a  $\Lambda$-model for elementary particles.  Qualitatively, following~\cite{Ze1}, we would propose a scenario of formation of a single direction of time evolution of microscopic substances, using the time-like vacuum potential. Namely, the global potential in 3D-time is able to generate strong quantum fluctuations in space-time. In particular, there are individual fluctuations being able to fix a time-like circular polarization in 3D-time equivalent to breaking of space-time symmetry and similar to the Higgs mechanism to induce the proper mass to a kind of identical elementary particles. The plane wave $(\ref{eq3})$ acquires a time-like circular polarization namely along the $t_3$ axis, keeping strictly an arrow evolution from the past to the future and being constrained by a time-like cylindrical condition. Polar coordinates $\{\psi(t_0),\varphi(t_0),t_3\}$ are used for the 3D-time instead of linear coordinates $\{t_k\}$ where vector $dt_0$ is evolving along a circle located in a plane orthogonal to the longitudinal vector $dt_3$. The squared differential along the time curve reads:

\begin{equation}
dt^2=d\psi(t_0)^2+\psi^2d\varphi(t_0)^2+dt_3^2=ds^2+dt_3^2;
\label{eq4}
\end{equation}

Where $ds$ is the Lorentz invariant interval characterizing the curvature term of time $t$, while the only linear term of time $dt_3$ in $(\ref{eq4})$ is getting identical to the time-like interval $d\vec{k}$ in  $(\ref{eq2})$, in according to the CLT principle. The direction of vector $dt_0$ serves a label attached to a particle or an anti-particle. Vectors $dt_3$ and $dt_0$ form a local basis for their vector summation $dt$ such as $\Omega dt=\Omega_0dt_0 + \Omega_3dt_3$, where $\Omega$, $\Omega_0$ and $\Omega_3$ are linear scale factorization parameters. Therefore, the local time-like geodesic $dt$ in 3D-time plays a role of a real time duration in our 3D-space.
\\Linear translation of a material point in 3D-space is taken along a linear vector $d\vec{l}$ defined along the same direction of the linear vector $d\vec{l}$ in $(\ref{eq2})$. Simultaneously, the material point can also rotate around an arbitrary 3D-spatial direction characterized by a spin $\vec{s}$. Originally, vector $\vec{s}$ in 3D-space is described by a spherical system attached to the material point $\{\psi(x_n),\theta(x_n),\varphi(x_n)\}$, where $\psi$ is the deviation variable extended in 3D-space, $\varphi$ is the azimuth around the spherical axis $\vec{l}$, while $\theta$ as the zenith between $\vec{l}$ and $\vec{\psi}$ defines its right-additive as the angle between $\vec{s}$ and $\vec{l}$. For conserving Lorentz invariance, we are restricted to consider the projection $s_n$ of $\vec{s}$ on axis $x_l$ which defines a local rotation in a plane $P_n$ orthogonal to vector $dx_l$ and the local proper coordinate $x_n\subset P_n$ now plays a role of a new affine parameter in 3D-space. The squared differential along the spatial curve reads:

\begin{equation}
d\lambda^2=d\psi(x_n)^2+\psi(x_n)^2\left [ d\theta^2+sin^2\theta d\varphi(x_n)^2\right]+dx_l^2=d\sigma^2+dl^2;
\label{eq5}
\end{equation}

Where $d{\sigma}^2=d\sigma_{ev}^2+d\sigma_{od}^2$ is a formal local interval in 3D-space; $d\sigma_{ev}$ is the P-even component related to rotation with the spatial symmetry and $d\sigma_{od}$ is the P-odd component implying a contribution of parity nonconservative (PNC) rotation. In general, the P-even term may contain a non-Lorentz invariant component of longitudinal fluctuations, however, the latter being compensated itself due to its forward-backward symmetry that does not contribute explicitly to our consideration.
\\In the extended time-space, $\psi=\psi(t_0,t_3,x_n,x_l)$ and $\varphi=\Omega_0t_0+\Omega_3t_3-k_n x_n-k_l x_l=\Omega t-k_jx_j;$ where $\{x_j\}$ are general spatial coordinates, describing both linear translation and rotation. Physical meaning of $\Omega$, $\Omega_0$ and $\Omega_3$ is defined as of time-like rotational velocities, while $k_j$, $k_n$ and $k_l$ are spatial wave factors. This procedure implies that two independent EDs $\psi$ and $\varphi$ turn into the dynamical parameters depending on other 4D space-time dimensions. Again, based on the CLT principle, the 6D "lightcone" geometry $(\ref{eq2})$ being generalized with curvature is turning into a new quadratic form:

\begin{equation}
dt^2-ds^2={dt_3}^2=dl^2=d\lambda^2-{d\sigma}^2;
\label{eq6}
\end{equation}

Accordingly, the 6D time-space representation is transformed to a 4D space-time geometry of a spinning particle describing both translation and rotation:

\begin{equation}
{d\Sigma}^2=ds^2-{d\sigma}^2=dt^2-d\lambda^2;
\label{eq7}
\end{equation}

Where $d\Sigma$ is a total 4D interval which is to be fixed as a Lorentz  invariant only after a transformation from the 6D time-space to the 4D space-time. The more generalized quadratic form $(\ref{eq7})$ includes, as its partial linearization case, the traditional quadratic form of 4D Minkowski space-time describing the linear translation:
  \begin{equation}
ds^2-d\sigma_{od}^2\approx ds^2=dt^2-dl^2;
\label{eq8}
\end{equation}

Where $d\sigma_{od}$ is too small as originated from PNC effect and usually may be ignored. Equation $(\ref{eq8})$ proves the consistency of geometry $(\ref{eq7})$ with special relativity.

\section{Geodesic deviation for gravitational wave}
\label{sec:Geodesic}

For formulation of the equation of motion in according to transformation from 6D time-space to 4D space-time we consider a geodesic acceleration equation of deviation $\psi$. The 3D local affine parameters in this equation are the transverse evolving time $t_0$ (proper time) and the transverse rotational space-like variables $x_n$ (proper spinning). In according to time-space symmetry $(\ref{eq2})$ we postulate that any deviation from the linear translation in 3D-time is to compensate by a deviation in 3D-space, i.e. a balance established between two extended differentials: $Du(t_0)=Du(x_n);$ where the velocity $u(s)=\frac{\partial \psi}{\partial s}$. Therefore, the symmetrical equation of geodesic acceleration of deviation $\psi$ in both 3D-time and 3D-space reads:

\begin{equation}
\frac{\partial^2 \psi}{\partial {t_0}^2}+\Gamma _{\alpha \beta }^{\psi }\left ( \frac{\partial t_\alpha }{\partial t_0} \right )\left ( \frac{\partial t_\beta }{\partial t_0} \right )=\frac{\partial^2 \psi}{\partial {x_n}^2}+\Gamma _{\gamma \sigma}^{\psi }\left ( \frac{\partial x_\gamma }{\partial x_n} \right )\left ( \frac{\partial x_\sigma }{\partial x_n} \right );
\label{eq9}
\end{equation}

Where $t_\alpha,t_\beta\in \{\psi(t_0),\varphi (t_0),t_3\}$ and $x_\gamma, x_\sigma \in \{\psi(x_n), \varphi(x_n),x_l \}; l,n=1\div3$. In both sides of $(\ref{eq9})$ among the related Christoffel symbols except  $\Gamma_{\varphi(t_0)\varphi(t_0)}^\psi=-\psi$  and $\Gamma_{\varphi(x_n) \varphi(x_n)}^\psi=-\psi.sin^2⁡\theta$  , all other terms are vanished or ignored due to orthogonality of $dt_\alpha$ (or $dt_\beta$) to $dt_0$ and of $dx_\gamma$ (or $dx_\sigma$) to $dx_n$. Due to the principle of CLT a differential equation of linear elements in according to  $(\ref{eq3})$, but with $\frac{\partial^2}{\partial t_3^2}$ instead of $\frac{\partial^2 }{\partial t_k^2}$, can be now added to  $(\ref{eq9})$ for fully describing the geodesic acceleration of deviation $\psi$ including rotation as well as linear translation:

\begin{equation}
\frac{\partial^2 \psi}{\partial {t_0}^2}- \psi\left ( \frac{\partial \varphi}{\partial t_0} \right )^2+\frac{\partial^2 \psi}{\partial {t_3}^2}=\frac{\partial^2 \psi}{\partial {x_n}^2}-\psi sin^2\theta\left ( \frac{\partial\varphi}{\partial x_n} \right )^2+\frac{\partial^2\psi }{\partial x_l^2};
\label{eq10}
\end{equation}

As in $(\ref{eq10})$  differentials $dt_3$ and $dt_0$, as well as corresponding covariant derivatives are locally orthogonal to each other, that we can group their second derivatives together as:

\begin{equation}
\frac{\partial^2 \psi}{\partial {{t_0}^+}^2}+\frac{\partial^2 \psi}{\partial {t_3}^2}=\frac{\partial^2 \psi}{\partial t^2};
\label{eq11}
\end{equation}

Where ${t_0}^+$ means an evolution toward the future. Similarly, due to a local orthogonality, for differentials $dx_l$ and $dx_n$, the second derivatives in 3D-space are also combined as:

\begin{equation}
\frac{\partial^2 \psi}{\partial {x_n}^2}+\frac{\partial^2 \psi}{\partial {x_l}^2}=\frac{\partial^2 \psi}{\partial {x_j}^2};
\label{eq12}
\end{equation}

In the result, two operations: i/ defining $\psi$ as a deviation parameter and ii/ the unification $(\ref{eq11})$ of two orthogonal time axes into the ordinary time $t$, almost hide the proper time $t_0$  and, simultaneously, reduce the 6D manifold into a 4D space-time. In principle, the vacuum circular polarization can separate time-like geodesic acceleration (as well as space-like one) in two opposite directions of evolution (or rotation). Before a separation of polarization there are symmetrical contributions of forward and backward evolutions in 3D-time and there is a symmetry between left-handed and right-handed contributions in 3D-space. However, due to transformation into 4D space-time and fixing the polarization, this symmetry is to be broken. For running the mechanism, qualitatively assuming that a Higgs-like tachyonic potential $V_H(\phi)=\lambda^2 \left [\phi^2-{\phi_0}^2\right ]^2$ induces a time-like centripetal force in 3D-time, where $\phi_0$ is the global vacuum field and $\lambda$ is the interaction constant. A centrifugal interaction potential in 3D-space is equivalent to a tachyonic centripetal in 3D-time and vise-versa which would convert to each other by spontaneous breaking of symmetry as: $V_H (\phi) \Rightarrow V_H (\chi)$ replacing $\phi=\chi+\phi_0$. This procedure is accompanied by transformations $(\ref{eq11})$ and $(\ref{eq12})$, simultaneously. Applying a dimensional analysis in natural units: $[length]=[time]=[mass]^{-1}$, we estimate the major term of acceleration of the time-like deviation  $\psi$ in $(\ref{eq10})$ as following:

\begin{equation}
-\left ( \frac{\partial \varphi}{\partial t_0} \right )^2\psi=\frac{V_H(\phi)}{\psi^2}\psi\Rightarrow -\left ( \frac{\partial \varphi}{\partial {t_0}^+} \right )^2\psi=\frac{V_H(\chi)}{\psi^2}\psi= C_1{m_H}^2\frac{\chi ^2}{\psi ^2}\left ( 1+\frac{\chi }{2\phi_0 } \right )^2\psi;
\label{eq13}
\end{equation}

Where the arrow means the breaking symmetry which is equivalent to selecting a single evolution toward the future accompanied by transformation from 6D time-space to 4D space-time. We propose that $(\ref{eq13})$ defines the major term of particle mass $m_0$ with a calibration factor of dimension $[C_1] = c^4\hbar^{-2}$. A square ratio of time-like radii can be used for a qualitative estimation: $\frac{\chi ^{2}}{\psi ^{^{2}}}\approx \frac{\left \langle \chi \mid r_H^2\mid \chi  \right \rangle}{\left \langle \psi \mid r_e^2\mid \psi  \right \rangle}=\left ( \frac{m_0}{m_H} \right )^{2};$ where $\left \langle r_H^2 \right \rangle$ and $\left \langle r_e^2 \right \rangle$ are square averaged classical radii of Higgs boson and the interacting elementary particle $e$, respectively.
\\The proper mass contains also space-like contribution corresponding to spinning in 3D-space which consists of P-even and P-odd terms. The squared P-even contribution is $m_s^2\sim \left ( k_n. \vec{s} \right )_{even}^2$, where $\vec{s}$ is the intrinsic spin of elementary particle and its normally oriented projection $s_l$ relative to the rotational plane $P_n$ can be both, left or right direction. In principle, electromagnetic or nuclear forces conserving the P-symmetry in 3D-space can be an inducing mechanism of the P-even contribution. Generally, without special polarization tool, the P-even contribution is to be hidden under the local geodesic acceleration condition in 3D-space:
\begin{equation}
\frac{\partial^2 \psi}{\partial {x_n}^2}-\psi sin^2\theta\left ( \frac{\partial \varphi}{\partial x_n} \right )^2\approx \frac{\partial^2 \psi}{\partial {x_n}^2}-\psi\left ( k_n.s_l \right )_{even}^2=0;
\label{eq14}
\end{equation}

The P-even mass contribution of the spinning term $m_s$ can be revealed in special polarization interactions. Instead of that, the P-odd contribution is a global proper polarization effect in term of the left-handed helicity, being observed universally in the weak interaction. We make a qualitative assumption that the breaking of symmetry in P-odd term may be caused by a vacuum potential of the global cosmological constant $\Lambda$ in 3D space:
\begin{equation}
\left ( \frac{\partial \varphi}{\partial x_n} \right )_{odd}^2\psi\Rightarrow \left ( \frac{\partial \varphi}{\partial x_n^L} \right )^2\psi=-\hbar^{-2}\omega^2s_L^2\psi;
\label{eq15}
\end{equation}

Where the arrow again means the breaking symmetry by fixing a given helicity; $s_L$ is a normal spin projection to the plane $P_n$ equivalent to the left-handed helicity. The mass scale factor $\omega$ of P-odd contribution can be estimated from electroweak interference of leptons as $\omega \sim \alpha.G_F m_0^3;$ i.e. proportional to the fine structure constant $\alpha$ and Fermi constant $G_F$.
\\Finally, from $(\ref{eq10})$ we obtain the 4D space-time geodesic equation as follows:

\begin{equation}
-\frac{\partial^2 \psi}{\partial t^2}+\frac{\partial^2 \psi}{\partial {x_j}^2}=-\left [ \left ( \frac{\partial \varphi}{\partial {t_0}^+} \right )^2-\left ( k_n.s_l\right )_{even}^2 -\left ( \frac{\partial \varphi}{\partial x_n^L} \right )^2 \right ]\psi=-{\delta}_M^2 \psi;
\label{eq16}
\end{equation}

This nonhomogeneous squared differential equation is a transmission law of the function $\psi$ which characterizes a time-like curvature with a space-like adjustment. Therefore, Equation $(\ref{eq16})$ defines the emission of microscopic gravitational waves from gravitational sources ${\delta}_M^2\psi$. At variance with the traditional gravitational waves with a very small curvature, Equation $(\ref{eq16})$ contains a major strong time-like curvature which is inversely proportional to the microscopic wave function $\psi$. Moreover, because the gravitational sources are globally extended everywhere in 6D time-space, the interaction potential is to attach to the moving material point, described by the geodesic equation $(\ref{eq16})$. In the result, the speed of translational transmission of the wave phase is faster than the speed of light, which seems to ban observation. Nevertheless, the energy-matter following Equation $(\ref{eq16})$ is to transmit with an observable subluminal speed. Indeed, going on to rescale $(\ref{eq16})$ with the Planck constant we can find that the microscopic gravitational waves are identical to the well-known and well-observable quantum waves.

\section{Quantum equations and indeterminism}

For formulation of quantum mechanical equations from a classical geodesic description, we adopt the quantum dynamic operators, such as: $\frac{\partial }{\partial t}\rightarrow i.\hbar\frac{\partial }{\partial t}=\widehat{E}$ and $\frac{\partial }{\partial x_j}\rightarrow -i.\hbar\frac{\partial }{\partial x_j}=\widehat{p_j}$. For the particle at rest, when $t\rightarrow t_0$ and $x_j\rightarrow x_n$, the operators are getting generators of proper masses: $i.\hbar\frac{\partial }{\partial t_0}=\widehat{E_0}=\widehat{m_0}$ and $-i.\hbar\frac{\partial }{\partial x_n}=\widehat{p_n}=\widehat{\delta m}$. This traditional procedure of quantum mechanics would be interpreted as a conversion between time and space which serves a mean for 4D space-time macroscopic observation of the microscopic gravitational waves transmitted in according to $(\ref{eq16})$.  Consequently, Equation $(\ref{eq16})$ leads to the basic quantum mechanical equation of motion:

\begin{equation}
-\hbar^2\frac{\partial^2 \psi}{\partial t^2}+\hbar^2\frac{\partial^2 \psi}{\partial x_j^2}-m^2\psi=0;
\label{eq17}
\end{equation}

Where the square mass term $m$ consists of the following components:   $m^2=m_0^2-\delta m^2=m_0^2-m_s^2-m_L^2$. In momentum representation Equation $(\ref{eq16})$ reads:

\begin{equation}
E^2\psi_p-\vec{p}^2\psi_p-m^2\psi_p=0;
\label{eq18}
\end{equation}

Equation $(\ref{eq18})$ describes subluminal motion of an elementary particle with energy $E$ and momentum $\vec{p}$. In comparison with the traditional expression of the rest mass, the present one includes an additional correction $\delta m$ associated with the contribution of the intrinsic spin in 3D-space. The P-even contribution $m_s$ linked with an external curvature of spinning in 3D-space can be compensated in according to the condition $(\ref{eq14})$ when only the linear translation along $x_l$ axis is taken in account for a laboratory frame observation. However, due to P-odd effect being observable in the weak interaction, the geodesic deviation of the material point by its spinning still induces a small non-zero mass scale factor $\left | \omega  \right |=m_L\ll m_s$ which proves a tiny internal curvature of our realistic 3D-space. In general, Equation $(\ref{eq17})$ is reminiscent of Proca equation of vector boson or the squared Dirac equation of lepton~\cite{Vo1}. In case of a scalar field or when there is no polarization analysis, $m \rightarrow m_0$, Equation $(\ref{eq17})$ turns to the traditional Klein-Gordon-Fock equation.
\\It is to mention that the local condition of geodesic deviation in 3D-space leads to:

\begin{equation}
\left ( \frac{\partial S}{\partial x_n} \right )^2=\left ( \hbar.sin\theta\frac{\partial \varphi}{\partial x_n} \right )^2=\frac{\hbar^2}{\psi}\frac{\partial^2 \psi}{\partial {x_n}^2}=-2mQ_B;
\label{eq19}
\end{equation}

Which is proportional to Bohm quantum potential $Q_B$ in ~\cite{Bo1}.
\\The existence of the spin term in $(\ref{eq17})$ is reminiscent of the Zitterbewegung of a free spinning electron (Schrodinger ZBW) ~\cite{Sc1}. In fact, when we describe in a laboratory frame a linear translation of a free particle by Equation $(\ref{eq17})$, the ZBW term is made almost hidden by the geodesic acceleration condition $(\ref{eq14})$ except a tiny P-odd term which is usually hard to observe. This implies a reason why ZBW is not observable experimentally without special measure of polarization or interference.
\\Suggesting that in the left side of $(\ref{eq9})$ the geodesic acceleration of deviation is restricted locally in 3D-time, one can derive the relation for time-energy indetermination:

\begin{equation}
\left | \Delta E \right |.\left | \Delta t \right |\geq \left | \Delta E_0 \right |.\left | \Delta t_0 \right |> \psi^{-1}\left | d\left ( i.\hbar\frac{\partial \psi}{\partial t_0} \right ) \right |.\left | dt_0 \right |=\left | i.\hbar \right |.d\varphi^2\geq \Delta \varphi_{min}^2\hbar\geq{0};
\label{eq20}
\end{equation}

Similarly, when the geodesic deviation acceleration is localized in 3D-space, the right side of $(\ref{eq9})$ leads to the space-momentum inequality:

\begin{equation}
\left | \Delta p \right |.\left | \Delta x \right |\geq \left | \Delta p_n \right |.\left | \Delta x_n \right |> \psi^{-1}\left | d\left ( i.\hbar\frac{\partial \psi}{\partial x_n} \right ) \right |.\left | dx_n \right |=\left | i.\hbar \right |.sin^2\theta d\varphi^2\geq \Delta \varphi_{min}^2\hbar\geq{0};
\label{eq21}
\end{equation}

Inequalities $(\ref{eq20})$ and $(\ref{eq21})$ can turn to equalities to zero only for the flat time-space of Euclidean geometry. For a non-zero curvature, it is proposed for $(\ref{eq21})$ an additional condition of  $sin^2\theta =1$, i.e. 3D-space quantization $\theta=(n+1/2)\pi$, equivalent to the cylindrical condition, and for both equations adopting $\Delta \varphi_{min}=\sigma\left(<\varphi>\right)=\sqrt{2\pi}$, where due to a statistical observability of the quantum indeterminism, $\sigma$ is a standard deviation of the mean value $<\varphi>=2\pi$ of an appropriate statistical distribution such as Poisson or Gaussian. In the result, we obtain the Heisenberg inequalities. This proves a direct link between internal time-space curvature of general relativity and Heisenberg inequalities in quantum mechanics.
Recalling that in ~\cite{Vo1} we derived the continuity equation of "a single particle" which in combination with the geodesic equation $(\ref{eq16})$ qualitatively allows to understand the physical reality of an individual particle in consistency with the quantum statistical interpretation and the context of wave-particle duality.

\section{Conclusions}

Based on a simplified model of time-space symmetry, we have shown that the extended 3D-time seems not to be a fictive subspace, but the existence of time-like EDs is revealed in terms of the quantum wave function $\psi$ and the proper time $t_0$ (under the time-like rotational parameter $\varphi$). Being hidden under curvature of the micro space-time, those two "transverse" time-like EDs contribute to a full time-space evolution of micro particles. A duality was found that the second order differential equations in quantum mechanics, including Proca, squared Dirac and Klein-Gordon-Fock equations, are indeed the emission law of microscopic gravitational waves which carry a strong time-like curvature in a weakly curved 3D-space.
\\The local 3D geodesic acceleration conditions of deviation $\psi$ shed light on the origin of quantum phenomena such as: i/ Bohm quantum potential, ii/ Schrodinger ZBW of a spinning electron and iii/ Heisenberg inequalities. In particular, triumph of Heisenberg indeterminism serves a strong evidence of internal curvatures of our realistic 4D space-time.
\\Those observations imply a necessity of reformation of our basic concepts, such as the flatness of 4D space-time and the locality of the dynamical interactions in 4D stage. It is possible to consider that the quantum mechanics is an effective 4D space-time hologram which serves for restoration of the physical world on a "lightcone" surface of the extended 6D symmetrical time-space.
\\Finally, the conclusions done here would serve positive arguments for a deep consistency between general relativity and quantum mechanics.

\section*{Acknowledgment}
The author is grateful to Ai Viet Nguyen and Anh Ky Nguyen (Hanoi Institute of Physics) for their useful discussion. A hearty thanks is extended to N.B.Nguyen (Hanoi Thang-Long Univ.) and M.L.Le-Vos for their technical assistance and precious encouragement.

\end{document}